\documentclass[preprint,12pt]{elsarticle}

\usepackage{amssymb}
\usepackage{amsmath}

\usepackage{xcolor}
\usepackage{soul}

\biboptions{sort&compress}

\journal{Chinese Journal of Physics}

\begin{document}

\begin{frontmatter}



\title{Denoising gravitational wave with deep learning in the time-frequency domain}


\author[label]{Yi-De Lee} 
\ead{yidelee@phys.ncku.edu.tw}

\author[label]{Hwei-Jang Yo\corref{cor1}} 
\ead{hjyo@phys.ncku.edu.tw}
\cortext[cor1]{Corresponding author}

\affiliation[label]{
            organization={Department of Physics, National Cheng Kung University},
            addressline={No.1, University Road}, 
            city={Tainan City},
            postcode={701}, 
            country={Taiwan}}

\begin{abstract}
Gravitational wave denoising is an ongoing task for revealing the events of compact binary objects in the universe.
Recently, with the aid of deep learning, gravitational waves have been efficiently and delicately extracted from the noisy data compared with the traditional match-filtering.
While most of the relevant studies adopt the data in the time series only, 
the time-frequency data processing is also in progress due to its several advantages for the waveform denoising. 
Here, we target the gravitational waves events emitted by binary black hole (BBH) mergers, with their total mass larger than 30 solar masses.
For denoising, we propose a deep learning model utilizing the Griffin-Lim algorithm, an existing numerical approach to restore the phase information from the related amplitude spectrogram.
This design allows extra attention on the phase recovery by using a priorly denoised amplitude spectrogram.
The denoising results fit well in both the amplitude and the phase alignments of the mock injected waveforms.
We also apply our model to the real detected events and discover a nice consistency with the simulated template waveforms, especially the high accuracy around the merger stage.
Our work suggests the possibility of a better methodological design for gravitational wave data analysis.
\end{abstract}

\begin{graphicalabstract}
\includegraphics[width=1.\textwidth]{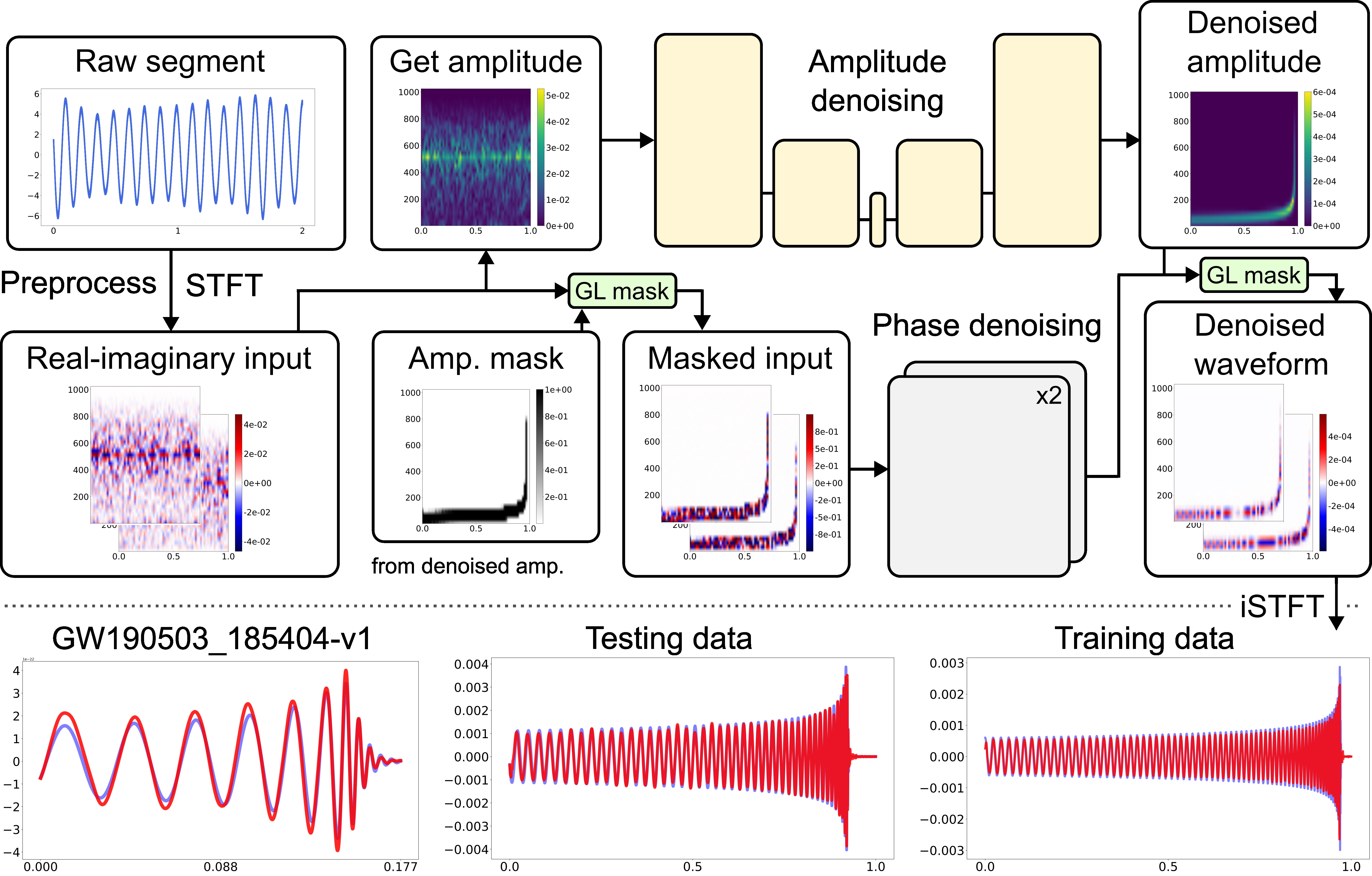}
\end{graphicalabstract}

\begin{highlights}
\item We propose a deep learning model for time-frequency gravitational wave denoising.
\item The Griffin-Lim algorithm inspires the design of the denoising workflow.
\item The amplitude is firstly denoised and it is used to simplify the phase recovery.
\item Our model denoises the gravitational waves with high accuracy.
\end{highlights}


\end{frontmatter}



\section{Introduction}
Benefiting from the detection of advanced LIGO \cite{collaboration2015advanced}, advanced VIRGO \cite{acernese2014advanced}, and other various interferometers \cite{luck2010upgrade, kagra2019kagra, unnikrishnan2013indigo}, gravitational wave is now a clear astrophysical phenomenon. 
Numerous studies have followed the discovery of gravitational waves, especially in the field of multi-messenger astrophysics \cite{abbott2017multi, meszaros2019multi, liu2021multi, corsi2024multi}.
While gravitational waves have been successfully detected, the waveforms are always immersed in the noises, which makes them invisible at first glance. 
To extract the waveform, the early studies relied on the match-filtering method \cite{allen2012findchirp}. 
By calculating the signal-to-noise ratio (SNR) between a segment of data and a simulated template, an event is considered to be identified if its SNR is above the false alarm level, and the template giving the highest SNR will be recorded as the best fit of this event. 
As the match-filtering method continues to make important progress in the gravitational wave detection and parameter estimation \cite{abbott2019gwtc, abbott2019binary, abbott2021gwtc, abbott2021population, abbott2023gwtc, abbott2023population}, it also requires a considerable number of waveform templates and a fair amount of computational effort to work \cite{indik2018reducing, dhurkunde2022hierarchical}.
As an alternative, the technique of deep learning has been applied to gravitational wave analysis because it is relatively efficient \cite{george2018deep, murali2023detecting} and capable of interpolating between the given templates \cite{george2018paraest}. 
Originated from the 1950s, deep learning has become a high-profile trend for a wide range of applications. 
The gravitational wave science, including detection \cite{george2018deep, george2018paraest, li2020some, krastev2021detection, mishra2021optimization, jiang2022convolutional, ma2022ensemble, nousi2023deep, qiu2023deep, jadhav2023towards, zhao2023space, bini2023autoencoder, chatterjee2024pre}, classification \cite{razzano2018image, krastev2021detection, lopac2021detection, murali2023detecting, qiu2023deep, sasaoka2024comparative}, noise reduction \cite{shen2019denoising, wei2020gravitational, chatterjee2021extraction, kato2022validation, 
bacon2023denoising, ghalsasi2023amplifying, jadhav2023towards, zhao2023space, murali2023detecting, wang2024waveformer, ma2025extraction}, and parameter estimation \cite{george2018paraest, green2020gravitational, gabbard2022bayesian}, greatly benefits from the rise of deep learning.

Since the quality of the data directly affects the performance of deep learning, the data preprocessing and representation are usually chosen to match the model structure of the deep neural network.
For the analysis of gravitational waves, data can be either expressed in the time domain or in the time-frequency domain, including their visualization.
In the early studies, many deep learning models worked with the (one-dimensional) time series instead of the (two-dimensional) time-frequency array.
Processing the time-series data is usually intuitive and could give the edge over storage consumption.
Furthermore, training the deep learning model with the time-frequency data may require more data processing and computational load due to the extra dimension. 
Despite these, the advantage of the time-frequency processing is that the frequency components are separated in the time-frequency domain, and thus the fluctuation covering the waveform is greatly reduced \cite{murali2023detecting}. 
Recently, Kato et al \cite{kato2022validation} designed a denoising deep learning model to capture the time-frequency gravitational waves with a non-harmonic analysis. 
Murali and Lumley \cite{murali2023detecting} inaugurated a deep learning autoencoder model to process the gravitational wave with its real and imaginary parts separated. 
Ghalsasi \cite{ghalsasi2023amplifying} showed that the deep learning denoising method performed well in recovering the amplitude of the time-frequency gravitational wave signals. 
Wang et al \cite{wang2024rapid} identified the time-frequency gravitational chirp from the LIGO noise with a deep learning network and presented the prediction in the form of the time-frequency spectrogram. 
These works indicate that it may be advantageous for the waveform denoising in the time-frequency domain. 

There are more benefits to suggest the feasibility of denoising with the time-frequency method.
For example, the behaviors of the noise and the waveform are more distinguishable in the time-frequency domain.
While various frequency components of the noise contribute to a continuous distortion in the time domain, they are unfolded in the time-frequency domain and become more or less discretely (and randomly) distributed.
On the contrary, the behavior of the gravitational waveform in the time-frequency domain is more or less regular and continuous.
Consequently, the discrimination between the waveform and the noise is expected to be more apparent before any further signal processing, although the magnitude of noise may be still (much) larger than the waveform (e.g., in Figure~\ref{fig:preprocess}).
Furthermore, the extra dimension of the time-frequency domain also allows various manipulations \cite{rioul1991wavelets, chatterji2005search, sejdic2009time, abbott2016observation, klimenko2016method, papandreou2018applications, cornish2020time, akan2021time, cornish2021bayeswave}.

\begin{figure*}[t]
    \centering
    \includegraphics[width=0.55\textwidth]{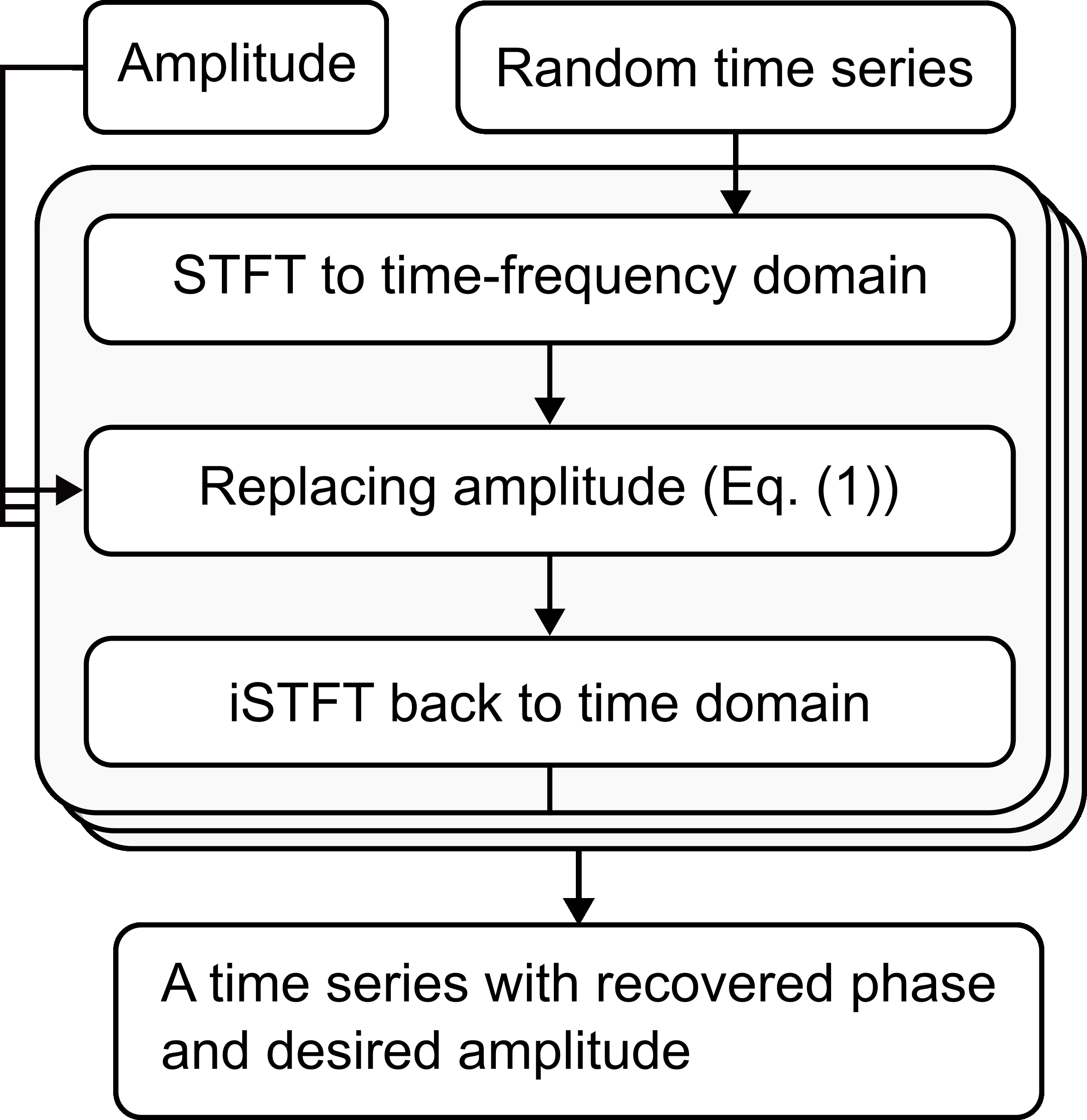}
    \caption{Visualizing the Griffin-Lim algorithm. The function of each block is described in the context. We note that Eq.~(\ref{glmask}) in this paper is the same as in the second block of the iteration.}
    \label{fig:gl}
\end{figure*}

Although the time-frequency method may be appealing, it could require some effort to preserve both the amplitude and phase to the end of the deep learning denoising process (see \cite{virtuoso2024wavelet} as a numerical example).
The amplitude of gravitational waves is continuous and potentially more visible than the noises in the time-frequency domain, while the phase evolution may be less obvious.
To our knowledge, Murali and Lumley  \cite{murali2023detecting} is the only research group that kept the phase well, while the other studies focused on the amplitude instead of the phase information \cite{kato2022validation, ghalsasi2023amplifying, jadhav2023towards, wang2024rapid}.
However, this shortcoming can be fixed if one begins with the clean amplitude denoised from a complex-valued time-frequency data. 
Here, we employ the Griffin-Lim algorithm \cite{griffin1984signal}, an iterative numerical method to estimate the corresponding phase evolution from the related amplitude solely, as the key to the denoising process. 
Utilizing the standard and the inverse short-time Fourier transformation (STFT), the Griffin-Lim algorithm aims to retrieve the phase information from a pure amplitude data (see Figure~\ref{fig:gl}, also \cite{liu2021novel}). 
It starts with a given amplitude spectrogram and a random time series, where the latter is used to initialize the phase information. 
The STFT firstly transforms the random time series into a complex-valued spectrogram in the time-frequency domain.
This random spectrogram is then normalized and rescaled by the given amplitude (see also Eq.~(\ref{glmask})), such that the phase is the only part that needs to be calibrated.
Next, the time-frequency spectrogram, with the desired amplitude and the random phase, is transformed back to the time domain by the inverse STFT, thus forming a complete loop of iteration.
By repeating the above-mentioned procedure, the transformation between the time domain and the time-frequency domain captures the phase overlap between the adjacent iterations, and the data converges to a single time series.
Therefore, with the Griffin-Lim algorithm, one can recover the phase information from a pure amplitude spectrogram.

Although the Griffin-Lim algorithm operates with the iteration method instead of the idea of deep learning, it provides the important hints about denoising the time-frequency gravitational waves.
Since the amplitude is repeatedly used for the phase recovery in the Griffin-Lim algorithm, one may also denoise the phase better with the aid of the cleaner amplitude. 
Fortunately, denoising the amplitude only is easier and has been performed with many deep learning models \cite{kato2022validation, ghalsasi2023amplifying, jadhav2023towards, wang2024rapid}, in which it showed that it is practical to obtain the clean amplitude in advance.
Meanwhile, it is worth mentioning that the Griffin-Lim algorithm assigns the given amplitude to the complex-valued spectrogram in every iteration, regardless of how the phase has been recovered.
The free combination of the amplitude and phase indicates that they can be separately processed, or even be thought of as two independent ingredients in the denoising workflow.
This allows some flexible and pioneering designs of the deep learning model architecture.
In addition, dividing the denoising task into the amplitude and phase steps in the time-frequency domain can justify the purpose of the corresponding model sub-structures, and enable us to precisely modify the parts for the improvement.
We then try to explore the potential of developing a time-frequency deep learning model for denoising the gravitational wave signal.

This work is organized as follows. 
In the next section, we explain the procedure of data preprocessing for the time-frequency waveform denoising, followed by the structure of the deep learning model inspired by the Griffin-Lim algorithm. 
We also provide the training strategy and demonstrate how we evaluate the denoising performance. 
In Section~\ref{sec:results}, we present the denoised waveforms of the mock injected events and the real cases for the BBH mergers. 
Finally, in Section~\ref{sec:discussion}, we discuss some future perspectives with a conclusion at the end.

\section{Methods}\label{sec:methods}
\subsection{Data preprocessing}\label{sec:dapre}

\begin{figure*}[t]
    \centering
    \includegraphics[width=1.\textwidth]{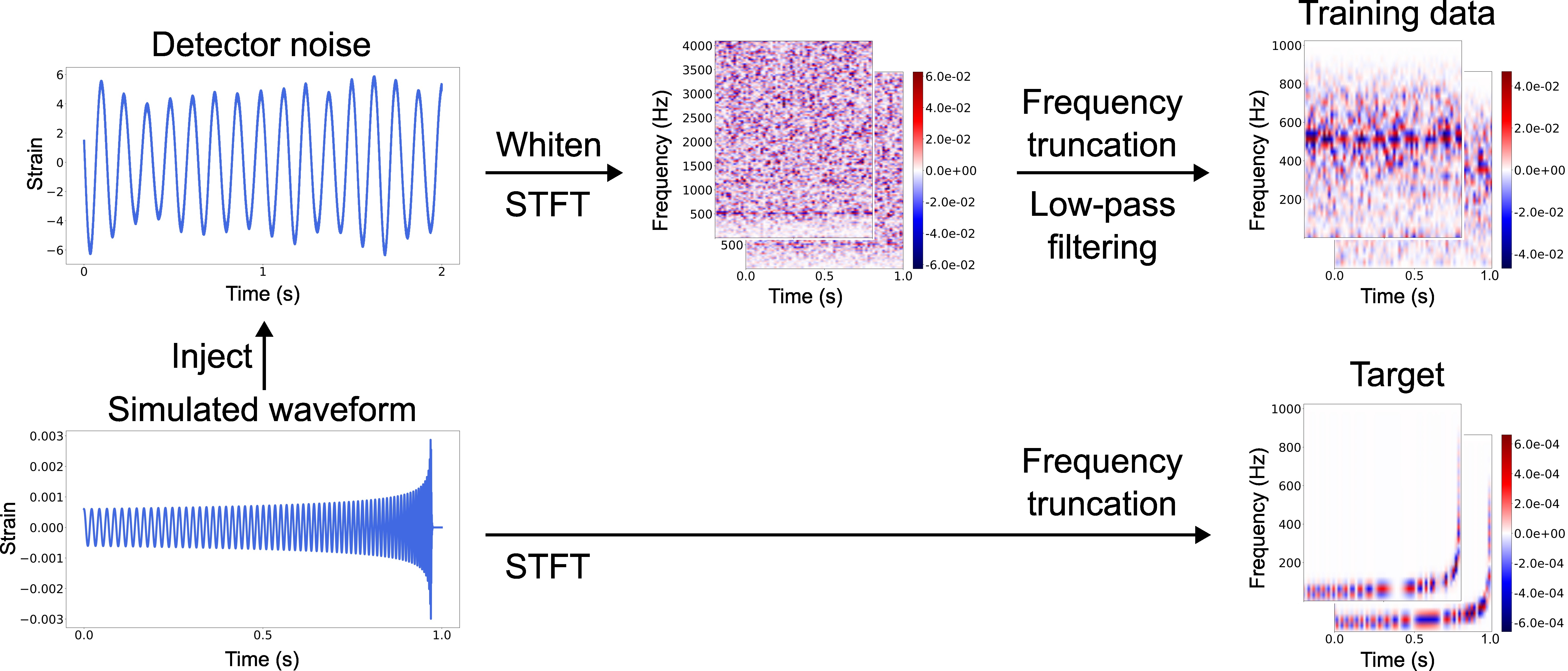}
    \caption{Data preprocessing for the time-frequency denoising.
    The procedure demonstrates how the noise and the waveform are combined into the training data and the target of the deep learning model. 
    The time-frequency panels consist of the real and the imaginary parts, which are plotted in the upper-left and lower-right, respectively.}
    \label{fig:preprocess}
\end{figure*}

To mimic the gravitational wave events, we inject the simulated waveform into a segment of the real LIGO detector noise. 
The noise is fetched from the Gravitational Wave Open Science Center (GWOSC) \cite{KAGRA:2023pio} and tailored into two-second length for each segment.
Only the data from the LIGO Hanford detector in the first stage of the third observation run (O3a) are used and we do not consider the cases with glitches in this study.
The waveform templates are generated by \texttt{SEOBNRv4} approximant \cite{bohe2017improved} in the \texttt{PyCBC} package \cite{alex_nitz_2024_10473621} with sampling frequency 8192 Hz.
We focus only on the BBH events by setting the mass parameters $m_{1}$ and $m_{2}$ ($m_{1}>m_{2}$) to the range from 5$M_{\odot}$ to 75$M_{\odot}$, and distribute these cases almost uniformly with some small perturbations.
As the counterpart of the luminosity distance, the SNR of the waveform covers the range $4\sim 12$, which is chosen according to the data of the real BBH merger events.
Technically, we manipulate the SNR through the scaling ratio of the normalized waveform within the range $0.001\sim0.045$ based on the relation described in \ref{app_rat_snr}.

We keep only the final second of the waveform segment including the merger stage, and smoothen the boundaries with a Tukey window \cite{harris2005use}. We then inject the one-second waveform into the two-second noise segment to obtain the mock event, where its merger stage is placed at the time of $1.2\sim 1.43$ second. 
This manipulation guarantees that every mock event segments contain the merger stage of a gravitational waveform in the subsequent processing, where the length ratio of the inspiral stage varies among the segments.
While it is possible to deal with the mock event segments containing only a part of the gravitational wave (e.g., \cite{wang2024waveformer}), we  prefer to adopt the waveform with all stages, i.e., the inspiral, the merger, and the ringdown stage (e.g., Figure~\ref{fig:preprocess}) in our first step.
In this work, we have to emphasize that we do not consider the cases with pure noise segments but only those having a single gravitational wave signal, since that these detected gravitational signals have been widely discussed \cite{george2018deep, george2018paraest, li2020some, krastev2021detection, mishra2021optimization, jiang2022convolutional, ma2022ensemble, nousi2023deep, qiu2023deep, jadhav2023towards, zhao2023space, bini2023autoencoder, chatterjee2024pre} and the purpose of this work is not for the detection but for the denoising.
To erase the colored frequency spectrum, the two-second mock event is whitened by the power spectral density (PSD) averaged from 5000 sets of the pure noise signal from the LIGO detection.
We then crop two 0.5-second segments at the beginning and at the end of each (two-second-length) signal to leave out the boundary effects from the whitening process. By doing so, the merger stage of the injected waveform is located roughly at $0.70\sim 0.93$ second.
The whitened mock event is also normalized by dividing the entire segment by its maximal value.

We apply the STFT to both the mock event and the simulated waveform to obtain their complex-valued time-frequency maps with 32 frequency bins and 64 time slices (see also \cite{dooney2025deepextractor}).
We implement the STFT due to its straightforward inverse transform \cite{dooney2025deepextractor} and the original usage in the Griffin-Lim algorithm. 
Although there are more delicate time-frequency transforms (e.g., the wavelet transforms \cite{rioul1991wavelets, chatterji2005search, virtuoso2024wavelet}) which can adjust the length of the time window for different frequencies and thus be better for the signal processing, it is easier for us to adopt the simple one in this work as our first step.
We apply a Hanning window \cite{harris2005use} to eliminate the frequencies larger than 512 Hz, so that the range of the BBH events can be accentuated. 
The training data for our deep learning model is the whitened-filtered mock event and our target data is the unwhitened simulated waveform. 
Here we emphasize that our model is trained to turn the noisy waveform back to the original untouched status in addition to the noise removal. 
Figure~\ref{fig:preprocess} illustrates the data preprocessing workflow.
\subsection{Deep learning architecture}\label{sec:deep}
\begin{figure}[!t]
    \centering
    \includegraphics[width=0.8\textwidth]{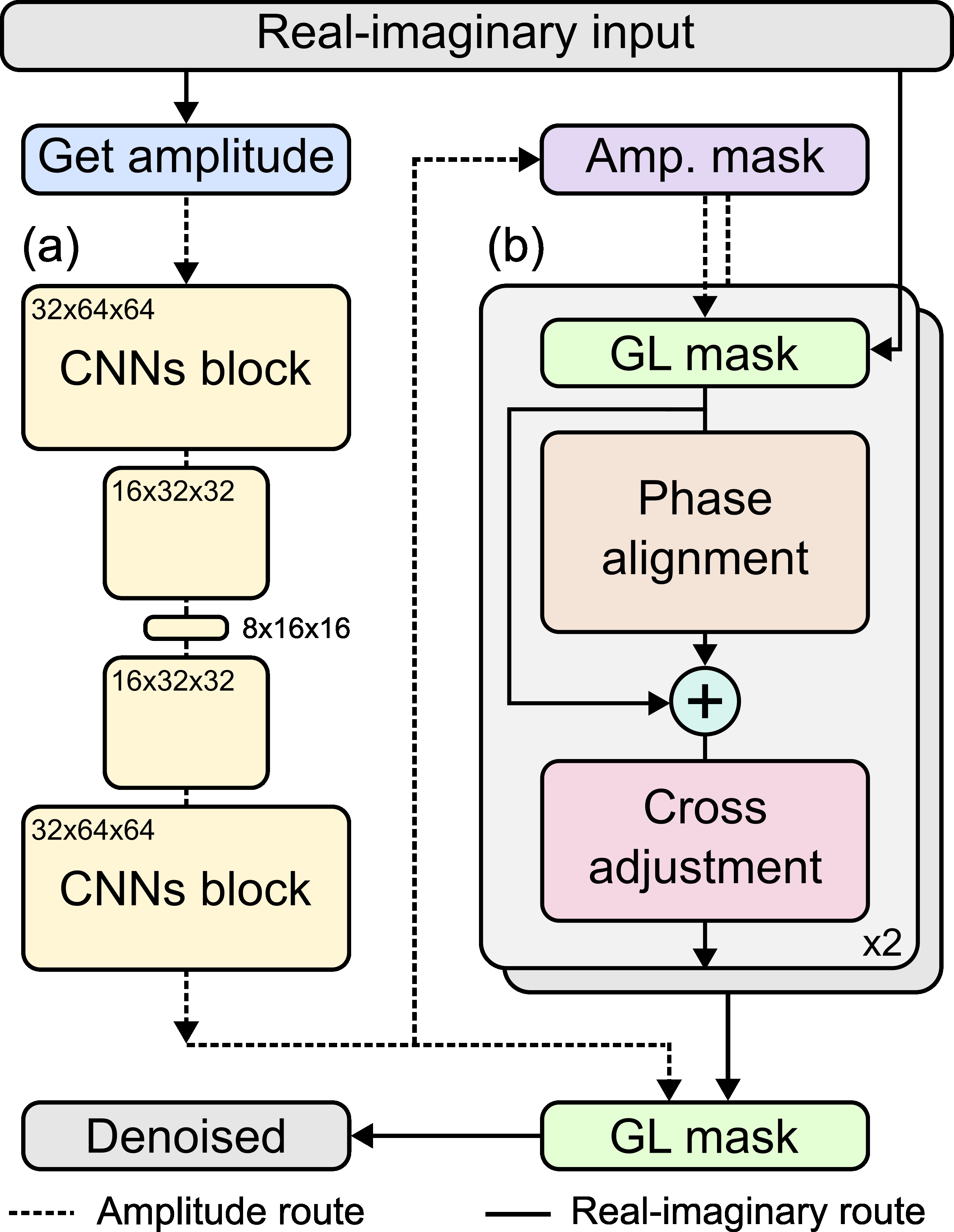}
    \caption{The denoising deep learning model architecture, which can be divided into (a) the amplitude autoencoder and (b) the Griffin-Lim network. The dashed and solid lines with arrows represent the amplitude and the real-imaginary data flow, respectively. Note that there are two inputs for the GL mask, one is the amplitude and the other is the real-imaginary data, as suggested in Eq.~(\ref{glmask}).}
    \label{fig:architecture}
\end{figure}
The denoising deep learning model consists of the amplitude autoencoder and the Griffin-Lim network. 
The amplitude spectrogram is firstly obtained with the amplitude autoencoder and the phase is then calibrated with the Griffin-Lim network.
Here, the amplitude is the magnitude of the complex data and the phase is carried by the real and the imaginary parts. 
Both parts of the model are based on the convolutional neural networks (CNNs) to extract the necessary features. 
The overall structure is organized as in Figure~\ref{fig:architecture}, while the further details are described in \ref{app_model_detail} and Figure~\ref{fig:app_model_detail}.
The cooperation of the two architectures allows retrieving the full information of the gravitational wave from the noisy signal in the time-frequency domain.

The amplitude autoencoder uses the structure of a typical autoencoder \cite{kramer1991nonlinear}. 
It is composed of a down-sampling encoder, a latent representation, and an up-sampling decoder. 
Figure~\ref{fig:architecture}(a) schematically displays its structure. 
Both the decoder and the encoder contain two CNNs blocks. 
In the network, the size of the time-frequency signal is compressed from (32, 64), (16, 32), to (8, 16) in the encoder and is recovered reversely in the decoder.
Meanwhile, the filter size of CNNs layers decreases from 64, 32, to 16 and then increases reversely. 
The down- and up-sampling are operated, respectively, by a two-dimensional convolutional and a two-dimensional transposed convolutional layer with a stride of 2 at the end and at the beginning of the CNNs blocks. 
Also, the convolution kernel with the size $7\times 7$ is dilated by the power of 2 to expand the local reception field. 
The {\tt Relu} activation function \cite{agarap2019deep} and the skip connections are installed within the blocks.
As an example, we sketch the first CNNs block of the amplitude autoencoder in Figure~\ref{fig:app_model_detail}(a) with more descriptions provided in \ref{app_model_detail}.
After the training, the amplitude autoencoder returns a clean amplitude spectrogram to be used for the upcoming phase reconstruction. 
Unlike the random phase initialization in the original Griffin-Lim algorithm, we now begin with the phase of the contaminated target, which is thought to be more precise than the random guess used in the Griffin-Lim algorithm.

The Griffin-Lim network features the flexibility of arbitrarily combining the amplitude and the phase for a time-frequency data and simplifies the phase denoising with respect to several aspects. 
For the application of the Griffin-Lim algorithm, we assign an amplitude spectrogram $A$ to a complex-valued spectrogram $X$ by an element-wise operation (also used in \cite{liu2021novel}):
\begin{equation}\label{glmask}
    \bar{X}=A\frac{X}{|X+\delta|}
\end{equation}
where $|\cdot|$ and $\bar{X}$ denote the amplitude and the new complex-valued spectrogram of $X$, respectively.
$\delta$ is an auxiliary parameter for stabilizing the division with small denominators and we set $\delta=10^{-8}$ here.
The purpose of Eq.~(1) is to replace the original amplitude of a complex-valued spectrogram $X$, which carries the phase information, with a specific amplitude $A$ such that the phase recovery can be achieved.
For the convenience, we name Eq.~(\ref{glmask}) as the Griffin-Lim mask (the GL mask in Figure~\ref{fig:architecture}(b)).
During the phase denoising, we obtain $A$ from the output of the amplitude autoencoder $D$ (Figure~\ref{fig:flow}(c)) with an analytically approximated element-wise Heaviside step function (see Figure~\ref{fig:flow}(d)):
\begin{equation}\label{stepfunc}
    A=\frac{1}{1+\mathrm{Exp}\left[-2k\left(D-\epsilon\right)\right]}
\end{equation}
where the parameters are set as $k=10^{5}$ and $\epsilon=5\times10^{-5}$. 
Eq~(\ref{stepfunc}) is in fact a sigmoid function.
None of the parameters in Eq.~(\ref{glmask}) and Eq.~(\ref{stepfunc}) are trainable and their values are chosen empirically. 
Provided that the non-trivial parts of the amplitude spectrograms tend to be greater than $10^{-4}$ due to the whitening process and the normalization during data preprocessing, it is legit to use these small values of $\delta$ and $\epsilon$.
We name Eq.~(\ref{stepfunc}) as the amplitude mask, and its purpose will soon be discussed in the next paragraph.
From the deep learning point of view, Eq.~(\ref{glmask}) and Eq.~(\ref{stepfunc}) can be thought of as the effective activation functions to fit our needs. 
We note that implementing the functions originating from numerical methods has been reported to be useful in reducing the number of training parameters and serving the purpose of the model \cite{masuyama2020deep}.

To illustrate the advantages of Eq.~(\ref{glmask}) and Eq.~(\ref{stepfunc}), we firstly apply the Griffin-Lim mask (Eq.~(\ref{glmask})) to the noisy input phase and the amplitude mask (Eq.~(\ref{stepfunc})), as shown in Figure~\ref{fig:flow}(e). 
In order to directly focus on the region of interest, the amplitude mask makes the magnitude of one uniformly inside and zero outside the contour of the waveform in the real-imaginary spectrogram. 
We treat all stages of the compact binary coalescence with equal significance during the phase reconstruction, regardless of the larger amplitude during the final merger or the smaller one during the earlier inspiral stage. 
Here we would like to emphasize that this kind of manipulation is hard to be operated in the time domain alone. 
This masking design is similar to the filter employed in Kato et al \cite{kato2022validation} but with more consideration on the waveform profile.

For the phase denoising, we adopt a structure similar to that of the CNNs blocks in the amplitude autoencoder to separately align the real and the imaginary phases but without any up- or down-sampling.
A cross-adjustment block is attached at the end of each iteration, where the real and the imaginary parts are interactively modified. 
The {\tt Selu} activation function \cite{klambauer2017self} is used in the Griffin-Lim network.
The detailed structures of the phase alignment block and the cross-adjustment block can be found in Figure~\ref{fig:app_model_detail}(b-c) and \ref{app_model_detail}.
After cross-adjusting, the amplitude mask is applied to the processed phase for another iteration. 
Eventually, the aligned phase is assigned to the denoised amplitude (i.e., $A=D$ in Eq.~(\ref{glmask}), as shown in Figure~\ref{fig:flow}(f)) to form the final output of the entire model. 
Note that we only borrow the idea of separating and recombining the amplitude and the phase from the Griffin-Lim algorithm \cite{griffin1984signal} without using any numerical iteration.
Also, the Griffin-Lim network used in this work is different and shall not be confused with the deep learning version of the Griffin-Lim algorithm (e.g., \cite{masuyama2020deep}).
\subsection{Training strategy}
For the purpose of the model's convergence, we apply several training strategies to our deep learning model. 
Firstly, we adopt the transfer learning to our training (see also \cite{george2018paraest, murali2023detecting}), where the amplitude autoencoder is trained before connecting to the Griffin-Lim network.
Secondly, with some experiments, we find that the curriculum learning is useful for finding a weak waveform covered by the overwhelming noise (see also \cite{george2018paraest}). 
To gradually increase the weight of the noise in the training data (i.e., the difficulty of the denoising task), the range of the scaling ratio is decreased from $0.06\sim0.08$ to the value in the normal BBH events (i.e., $0.001\sim0.045$) in the initial 100 epochs and henceforth remains unchanged for another 300 epochs. 
Once the amplitude autoencoder converges, we freeze the parameters and turn to train the Griffin-Lim network with the same data set described in Section~\ref{sec:dapre} for 400 epochs.

In total, 80000 noise segments and 90600 waveforms are collected for the training.
Only 30000 injections are produced in each epoch by changing the amplitude of the waveforms and choosing the noise segments randomly.
Owing to the transfer learning and the curriculum learning we use, changing the training data every epoch will not prevent the model from converging.
Instead, we find that such strategy effectively diversifies the data and thus avoids overfitting.
The model parameters are calibrated by minimizing the difference between the output and the clean target waveform, practically realized by reducing the root-mean-square error (RMSE). 
For the Griffin-Lim network, the total loss is the summation of the real and the imaginary loss with the amplitude mask (similar to \cite{wang2024waveformer}). 
The training is optimized with the Adam algorithm \cite{kingma2014adam}. 
The learning rates of the two trainings are both adapted to be smaller when the values of the loss functions start to converge. 
There are 7,831,011 trainable parameters in the entire model, where 2,198,225 parameters and 5,632,786 parameters belong to the amplitude autoencoder and the Griffin-Lim network, respectively. 
This model is trained on 8 NVIDIA Tesla V100 GPUs with 32 cores using TensorFlow library \cite{abadi2016tensorflow} for about a day.
\subsection{Performance evaluation}\label{sec:perf}
Once the denoised waveform is obtained, we pad the truncated frequency parts (i.e., $1024\sim 4096$ Hz, see Section~\ref{sec:dapre}) with zeros and then transform it back to the time domain with the inverse STFT for visualization. 
The overlap \cite{abbott2021gwtc} between the denoised waveform $h$ and the target waveform $s$ can be calculated with
\begin{equation}
\mathcal{O}(s,\,h)=(\hat{s}\,|\,\hat{h}),
\end{equation}
where
\begin{equation}
(s\,|\,h)=2\int_{f_{\rm min}}^{f_{\rm max}}\frac{\tilde{s}(f)
\tilde{h}^*(f)+\tilde{s}^*(f)\tilde{h}(f)}{S_n(f)}{\rm d}f,
\end{equation}
is the noise-weighted scalar product \cite{finn1992detection},
$\hat{h}=h/\sqrt{(h\,|\,h)}$, 
$\hat{s}=s/\sqrt{(s\,|\,s)}$
are the normalized time series, and $S_{n}(f)$ is the PSD used for whitening.
$f_{\rm min}$ and $f_{\rm max}$ are taken to be 0 and 4096 Hz, respectively. 
The tilde notation denotes the Fourier components of the corresponding time series.

We prepare a test set of 14,399 mock events, not used in the training, with the same mass pattern as in the training data set (Section~\ref{sec:dapre}).
The mass uniformly covers the range $4.5\sim 82.6 M_{\odot}$ with small perturbations (Figure~\ref{fig:statistics}(a)). 
The events with small SNRs account for a larger population in the range $4\sim 25$ (Figure~\ref{fig:statistics}(b)).
Specifically, about 74\% of the SNR values are less than 10, which is the median of the recorded SNRs of the BBH mergers.
In addition, we focus only on the mock events with $m_{1}+m_{2}\gtrsim30 M_{\odot}$ (i.e., medium-small or larger total masses), because BBH mergers with small total masses can result in high frequency waves and may be easily confused with the transient noise.
While we only consider a portion of the possible range of the total mass, those cases still account for 80$\%$ of the real BBH mergers recorded by GWOSC until the latest event GW250114\_082203 \cite{abac2025gw25011}.
As to the real BBH mergers, we collect the detected events from LIGO Hanford, and simulate the template waveforms with \texttt{SEOBNRv4\_opt}~\cite{devine2016optimizing}, \texttt{SEOBNRv4PHM}~\cite{ossokine2020multipolar} or \texttt{SEOBNRv5PHM}~\cite{ramos2023next} approximant.
\begin{figure*}[!htbp]
    \centering
    \includegraphics[width=1.\textwidth]{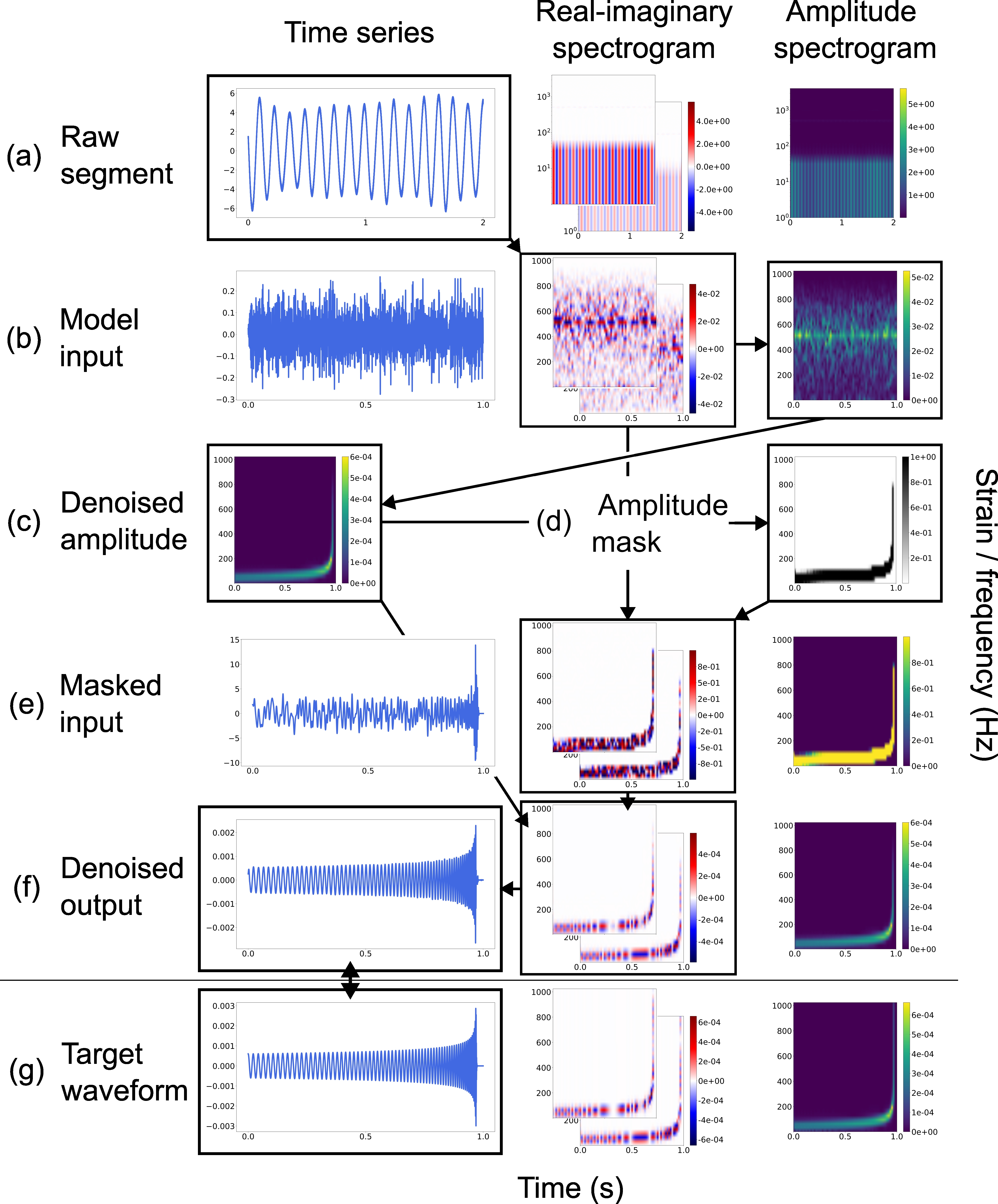}
    \caption{The workflow of denoising an injected gravitational wave event. We specify the important stages, (a) raw detector data segment with a gravitational wave signal, (b) preprocessed data, (c) the denoised amplitude spectrogram, (d) the binary amplitude mask calculated from (c), (e) the model input after begin masked by (e), (f) the denoised waveform, and (g) the target waveform. The frames of the panels indicate the current processing representation. The arrows between the panels point out the direction of the data flow. The real and the imaginary parts are respectively placed on the upper-left and the lower-right in the second column. There are no time series or real-imaginary spectrograms of stage (c) and (d). Only the frequency axes of the time-frequency spectrograms of (a) are plotted in the logarithm scale for a better visualization.}
    \label{fig:flow}
\end{figure*}
\section{Results}\label{sec:results}
\subsection{Dataflow of the denoising procedure}\label{subsec:flow}
We summarize the denoising workflow in Figure~\ref{fig:flow}.
One can see that the raw data contains large magnitudes of low-frequency noise (Figure~\ref{fig:flow}(a)).
After some signal processing, the magnitudes of different frequency components are redistributed and a vague chirp is visible in the lower-right corner of the amplitude spectrogram (Figure~\ref{fig:flow}(b)). 
The amplitude spectrogram is firstly denoised (Figure~\ref{fig:flow}(c)) and then used to generate the amplitude mask (Figure~\ref{fig:flow}(d)), so that the attention could be drawn into the non-trivial waveform region (Figure~\ref{fig:flow}(e)).
One can even only apply the amplitude mask in the process to obtain a rough shape of the gravitational wave as in the time domain (Figure~\ref{fig:flow}(e)).
The phase recovery then starts with the masked real-imaginary data (Figure~\ref{fig:flow}(e)) and ends with the recombination of the aligned phase and the denoised amplitude (Figure~\ref{fig:flow}(f)).
We notice that although the zigzags along the contour of the amplitude mask (Figure~\ref{fig:flow}(d)) may include some unwanted region during the phase recovery (Figure~\ref{fig:flow}(e)), the final assignment of the denoised amplitude automatically ignores such defect and produces a smooth envelope for the output waveform (Figure~\ref{fig:flow}(f)).
Compared with the target waveform (Figure~\ref{fig:flow}(g)), the model presents the ability to capture the essence of the gravitational waves from the noisy data.
\subsection{Denoising mock events}\label{sec:mock}
\begin{figure*}[!t]
    \centering
    \includegraphics[width=1.\textwidth]{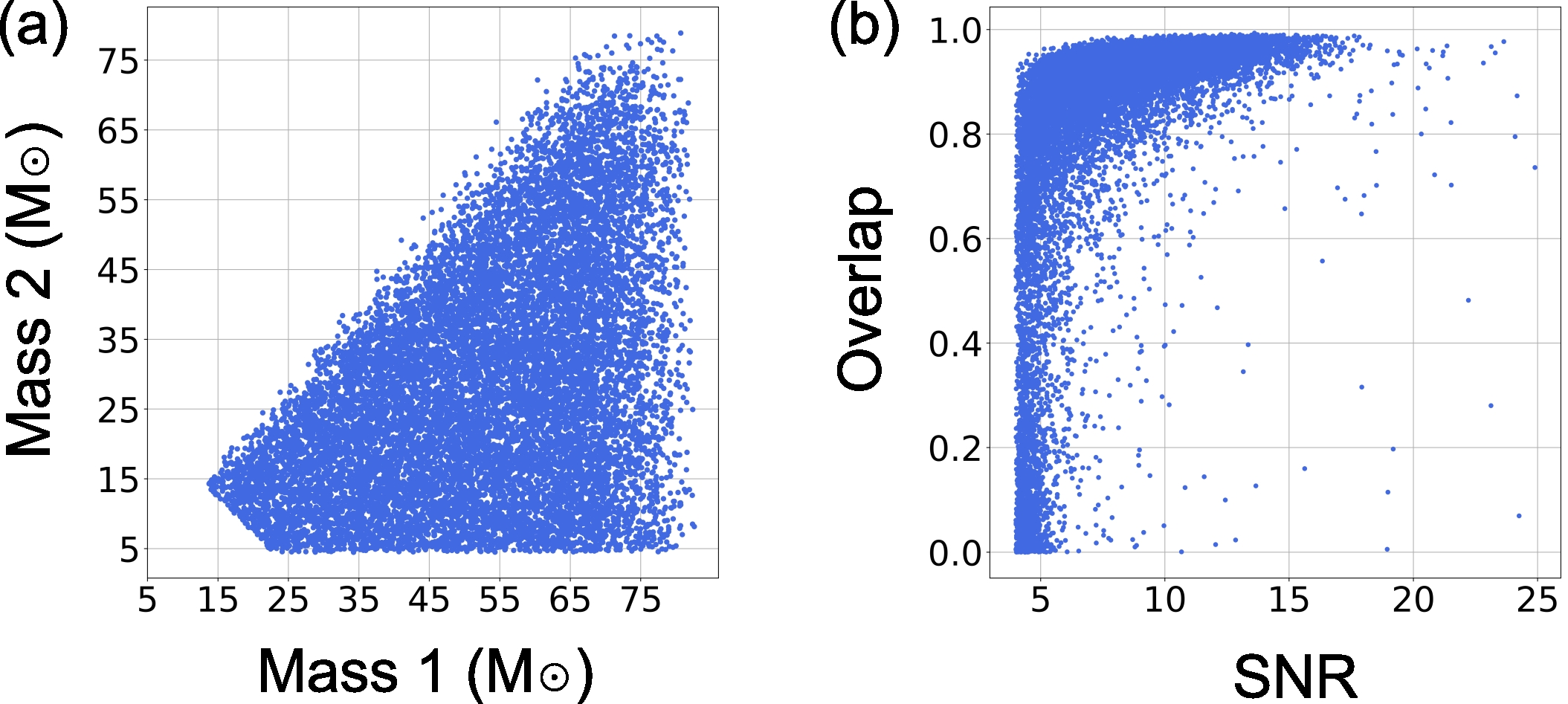}
    \caption{Statistics of the 14,399 denoised injected mock events. (a) Distribution of the mass parameters of the mock test set. The masses are in the unit of solar mass. The missing population in the lower-left corner is due to the exclusion of the events with small total masses. (b) Relation between the SNR and the overlap. It can be observed that large SNR is more likely to result in large overlap.}
    \label{fig:statistics}
\end{figure*}
\begin{figure*}[!t]
    \centering
    \includegraphics[width=1.\textwidth]{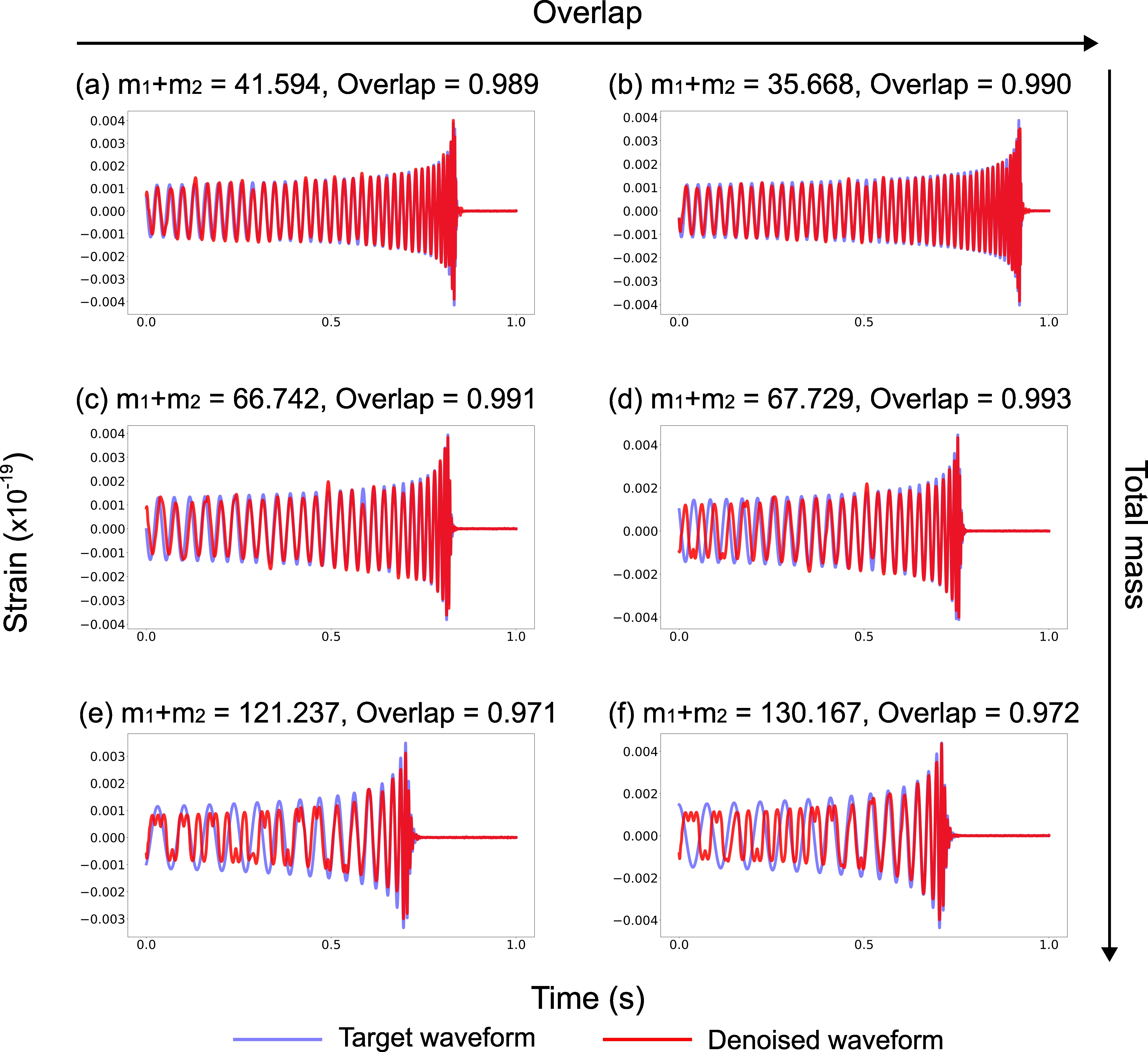}
    \caption{Comparison between the denoised mock events (red) and the target waveforms (blue) in the time domain. The overlap increases from left to right, while the total mass increases from top to bottom. It can be observed that overlaps of the events with medium-small (a-b) or medium total masses (c-d) are higher than those with large total masses (e-f). Also, the alignments between the denoised and the target waveform are generally better in the merger stage compared to the inspiral stage.}
    \label{fig:inject}
\end{figure*}
In this work, we test the model by the mock events where the masses are uniformly distributed with small perturbations (Figure~\ref{fig:statistics}(a)), and show that it is able to obtain high accuracy in denoising (Figure~\ref{fig:statistics}(b)).
Among the 14,399 test data sets, the overlaps of 10,681 denoised sets (74\%) with the related (clean) waveforms are greater than 0.8. 
In particular, the cases with large SNR, usually representing short luminosity distances, result in better reconstructions (Figure~\ref{fig:statistics}(b) and Figure~\ref{fig:rat_snr_olp} in \ref{app_rat_snr}).
It can be verified that among the high-overlap ($>0.8$) cases, 6,612 events (62\%) have their SNRs greater than 8.
Notably, some cases with smaller SNR can also lead to large overlap, though the portion is relatively low (Figure~\ref{fig:statistics}(b)).

Here we display several denoised mock events with high overlaps for medium-small, medium, and large total masses in Figure~\ref{fig:inject}.
The overlap turns out to be high for those cases with medium-small and medium total masses (Figure~\ref{fig:inject}(a-d)), while the mass ratio does not significantly affect the results.
Among the well-denoised waveforms, the model can accurately extract not only the merger stage but also the (late-)inspiral stage, representing the success of the model design on treating all stages with equal weighting.
Meanwhile, the early-inspiral stage tends to be less recovered, which should be expected since its behavior is more sensitive to the (non-mass) physical parameters than those during the merger stage.
Generally, we note that the denoising performance for events with larger total masses is not as good as for lighter events (Figure~\ref{fig:inject}(e-f)).
This may be the consequence of their relatively low frequencies, which may be partially suppressed during data preprocessing (Section~\ref{sec:dapre}). 
It deserves to investigate more in our following work.
\begin{figure*}[!t]
    \centering
    \includegraphics[width=1.\textwidth]{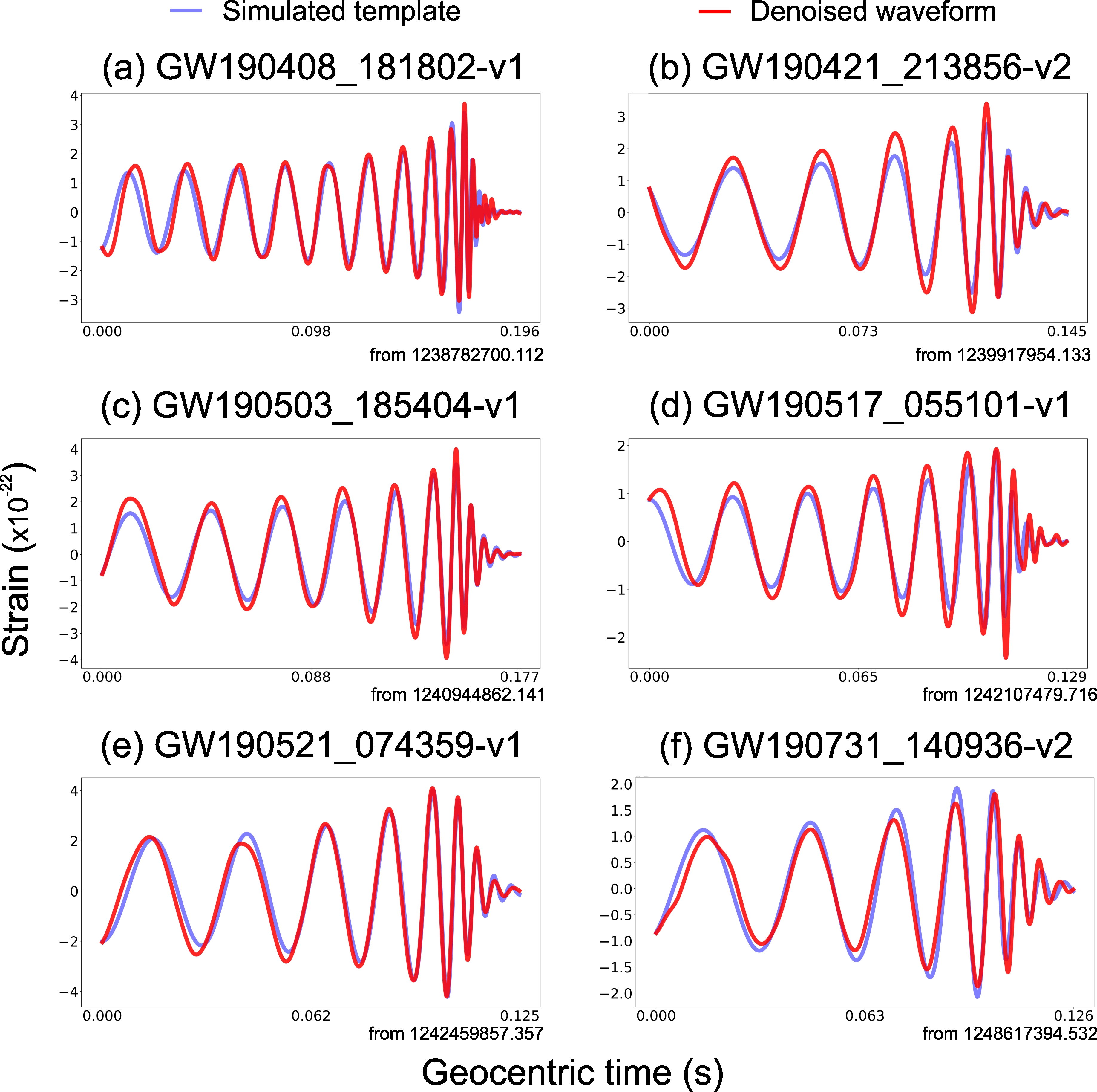}
    \caption{Comparison between the denoised BBH merger waveforms (red) and the simulated templates (blue) of the events during O3a with ideal match around the merger stage. All of the presented waveforms are not whitened or filtered. The denoised waveform and the simulated template coincide more in the merger stage while tend to deviate more in the inspiral stage.}
    \label{fig:o3a}
\end{figure*}
\begin{figure*}[!t]
    \centering
    \includegraphics[width=1.\textwidth]{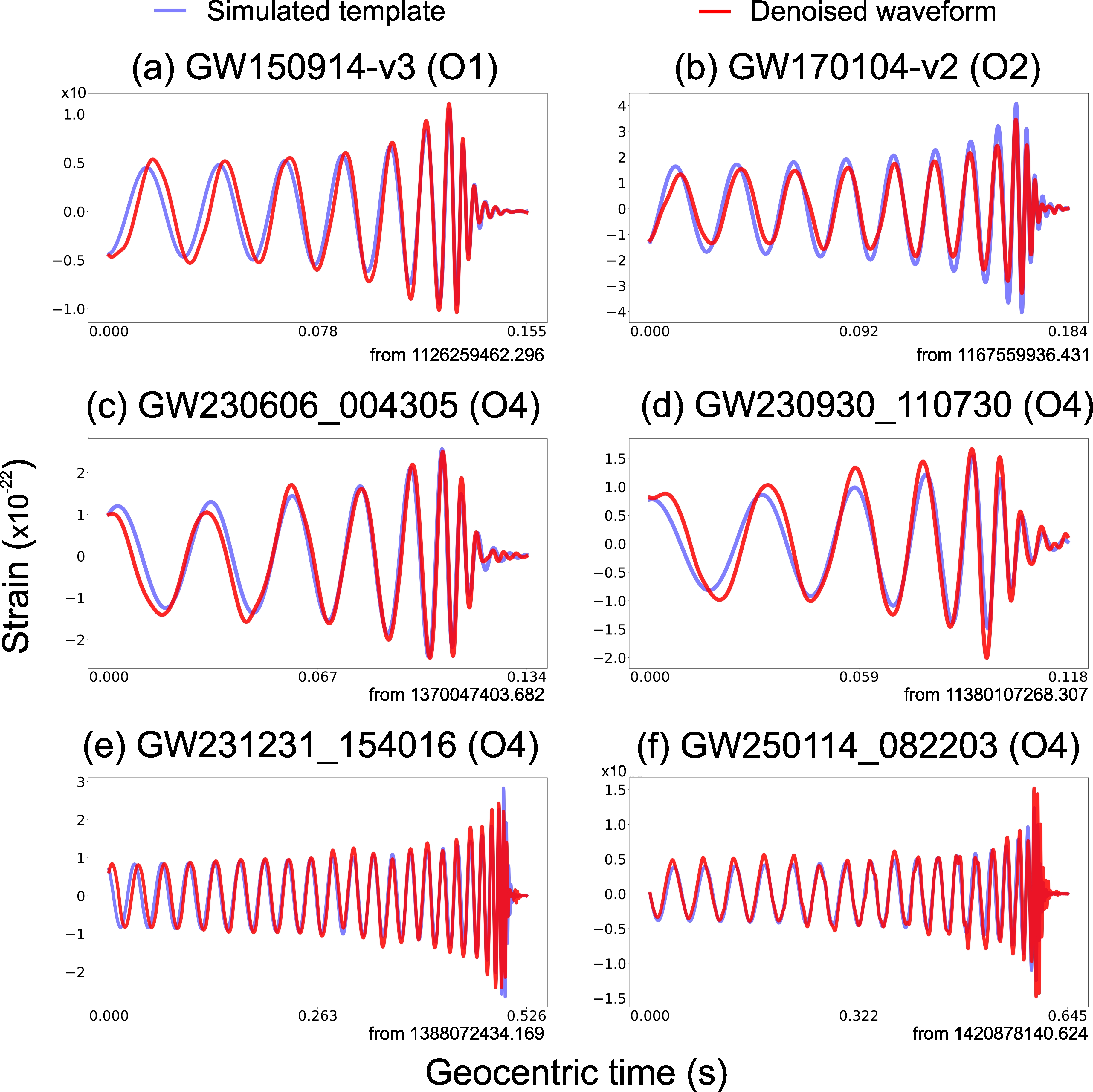}
    \caption{Comparison between the denoised BBH merger waveforms (red) and the simulated templates (blue) of the events during other observation runs with ideal match. 
    All of the presented waveforms are not whitened or filtered. 
    The strain values of the events GW150914-v3 (a) and GW250114\_082203 (f) are 10 times larger than others. 
    These events are arranged in the increasing order of their geocentric times. 
    (a-d) The denoised waveform and the simulated template coincide in the merger stage while tend to deviate more in the inspiral stage. 
    (e-f) The early-inspiral stages are also well-denoised.}
    \label{fig:others}
\end{figure*}
\subsection{Denoising real BBH merger events}
For the cases of the real BBH merger, we find that the result is acceptable and consistent for the comparison between the denoised waveforms and the simulated templates around the merger stage during O3a.
Some representative denoised events are shown in Figure~\ref{fig:o3a} with the suitable durations to highlight the fine alignments.
While the two waveforms agree rather well in the merger stage, they are likely to deviate during the early-inspiral stage, which has also been observed when denoising the mock events (Section~\ref{sec:mock}).

For curiosity, we apply the model trained with the noise during O3a to denoise the events recorded in the other periods, including the first (O1), the second (O2), and the fourth (O4) observation runs of LIGO Hanford (Figure~\ref{fig:others}).
Even though the PSDs of these runs are different from the one used in training, our model is still able to denoise the events in other observation runs, indicating that it has been well trained to distinguish the waveform from the noise for the LIGO Hanford detector. 
Similar to the results in Figure~\ref{fig:o3a}, the first four denoised data in Figure~\ref{fig:others}(a-d) are more aligned and consistent in the merger stage than in the inspiral stage.
Nevertheless, the early-inspiral stages of the events GW231231\_154016 and GW250114\_082203 (Figure~\ref{fig:others}(e-f)) are also recovered surprisingly well.
Notably, the events GW150914-v3 (Figure~\ref{fig:others}(a)) and GW250114\_082203 (Figure~\ref{fig:others}(f)) are the first and the latest detected events until now, suggesting that our model is possible to be applied throughout the observation.
These denoising results show that our model could have the potential to provide further details of the BBH merger events after their detections.

\section{\label{sec:discussion}Discussing conclusion}
We construct a deep learning model based on the Griffin-Lim algorithm to denoise the time-frequency gravitational waveforms (Figure~\ref{fig:architecture}). 
By separating the denoising process of the amplitude and phase, we reach promising accuracy in recovering the mock injected events with about three quarters of the overlaps exceeding 0.8 (Section~\ref{sec:mock} and Figure~\ref{fig:inject}).
Our model is also capable of capturing the merger stage of real events (Figure~\ref{fig:o3a}) with the mass information only. 
In order to discover more details of the detected waveforms, expanding the parameter space of training waveforms, such as including both spins of the black holes or the inclination, should be still necessary and helpful. 
Additional choices of waveform approximants, for instance \texttt{IMRPhenomD} \cite{husa2016frequency, khan2016frequency}, \texttt{SEOBNRv4\_opt} \cite{devine2016optimizing}, \texttt{IMRPhenomXPHM} \cite{pratten2021computationally}, or \texttt{SEOBNRv4PHM} \cite{ossokine2020multipolar}, may also diversify the data and possibly improve the alignments. 
To obtain a better recovery of the weak inspiral stage, prolonging the training segments should be a promising way to provide more information on the early signal, and to allow an in-depth denoising for the part of smaller wave amplitude.
Moreover, in the aspect of signal preprocessing, we note that the wavelet transforms \cite{rioul1991wavelets, chatterji2005search, virtuoso2024wavelet} do a nice job in presenting the data, and the denoising performance may be improved using the time-frequency map obtained from the wavelet transforms.
Nevertheless, a further investigation of the denoising model might be needed to handle more complicated data to give a better performance. 

Our model features the amplitude mask (Eq.~(\ref{stepfunc})) which prevents the denoising model from overemphasizing the merger stage.
It can be seen from Figure~\ref{fig:inject}(a-d) and Figure~\ref{fig:others}(e-f) that the inspiral stage is well recovered despite its smaller amplitude.
We point out that further modification of the model can be done by allowing the co-adjustment of amplitude and phase.
Since the phase recovery relies on the data masked by the denoised amplitude (Figure~\ref{fig:flow}(e)), it may be difficult to work if the denoised amplitude deviates too much from the expectation.
Although currently there is not yet a thoughtful strategy to interactively improve these two substructures, the design of the amplitude mask has taken the lead in dealing with the knotty problem of denoising the waveform, particularly for both the inspiral and the merger stages.

The key advantage of our denoising model is the free combination of the amplitude and the phase of the time-frequency data. 
With the amplitude mask (Figure~\ref{fig:flow}(d)), the complicated phase reconstruction can be eased by adjusting all the amplitudes to the same magnitude and getting rid of the unwanted parts (Figure~\ref{fig:flow}(e)).  
This kind of manipulation is similar to the attention mechanism \cite{vaswani2017attention} in deep learning, especially for the two-dimensional cases (e.g., \cite{dosovitskiy2020image}). 
Since the attention mechanism has reached astonishing achievements in many applications \cite{niu2021review}, it is interesting to employ the mechanism for the time-frequency data processing in the near future.

This study represents our first step in the combination of numerical analysis and deep learning for time-frequency gravitational wave processing. 
Numerical analysis formulates the algorithm cleverly with limited resources, while deep learning is known for the universal capability by the emergent behavior of numerous nodes and links \cite{cybenko1989approximation, hornik1989multilayer, leshno1993multilayer}. 
We underline that further improvements can be achieved in gravitational wave data processing or even in the pipeline design.
In the near future, we will try to explore the capability of this model with more complicated training data by adding the spin parameters of the BBH mergers, and also to improve the model's ability of denoising with more computational power.

\section*{Acknowledgements}
We thank Prof.~C.S.~Lue for sharing computational resources. 
We also thank Huy Gia Tong, Richard~Guh, and S.C.~Ho for their constructive comments on the manuscript. 
Our idea on the amplitude mask benefits from conversations in the competition of CTCI Outstanding Physics Undergraduate Award, held by the Physical Society of the Republic of China (PSROC) in Taiwan.
We acknowledge National Center for High-performance Computing (NCHC) for providing computational and storage resources. 
This research has made use of data or software obtained from the Gravitational Wave Open Science Center (gwosc.org), a service of the LIGO Scientific Collaboration, the Virgo Collaboration, and KAGRA.

\begin{figure}[!t]
    \centering
    \includegraphics[width=0.7\textwidth]{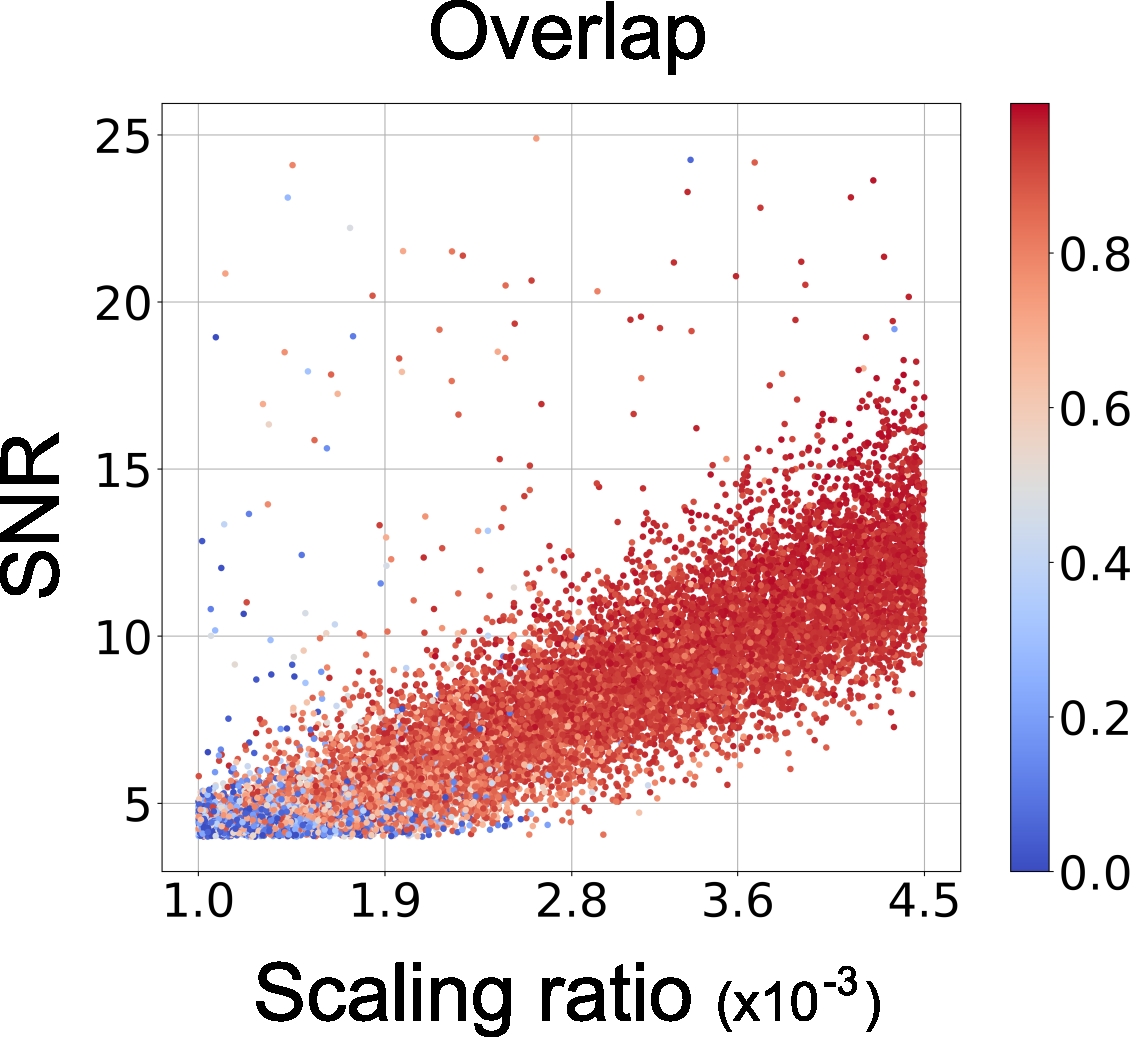}
    \caption{Relation between the scaling ratio, the SNR, and the overlap of the 14,399 injected mock events. The color gradient shows the values of the overlap in the scatter plot. The relation presents a clear trend that both the SNR and the overlap increase with the scaling ratio.}
    \label{fig:rat_snr_olp}
\end{figure}
\begin{figure}[!t]
    \centering
    \includegraphics[width=1.\textwidth]{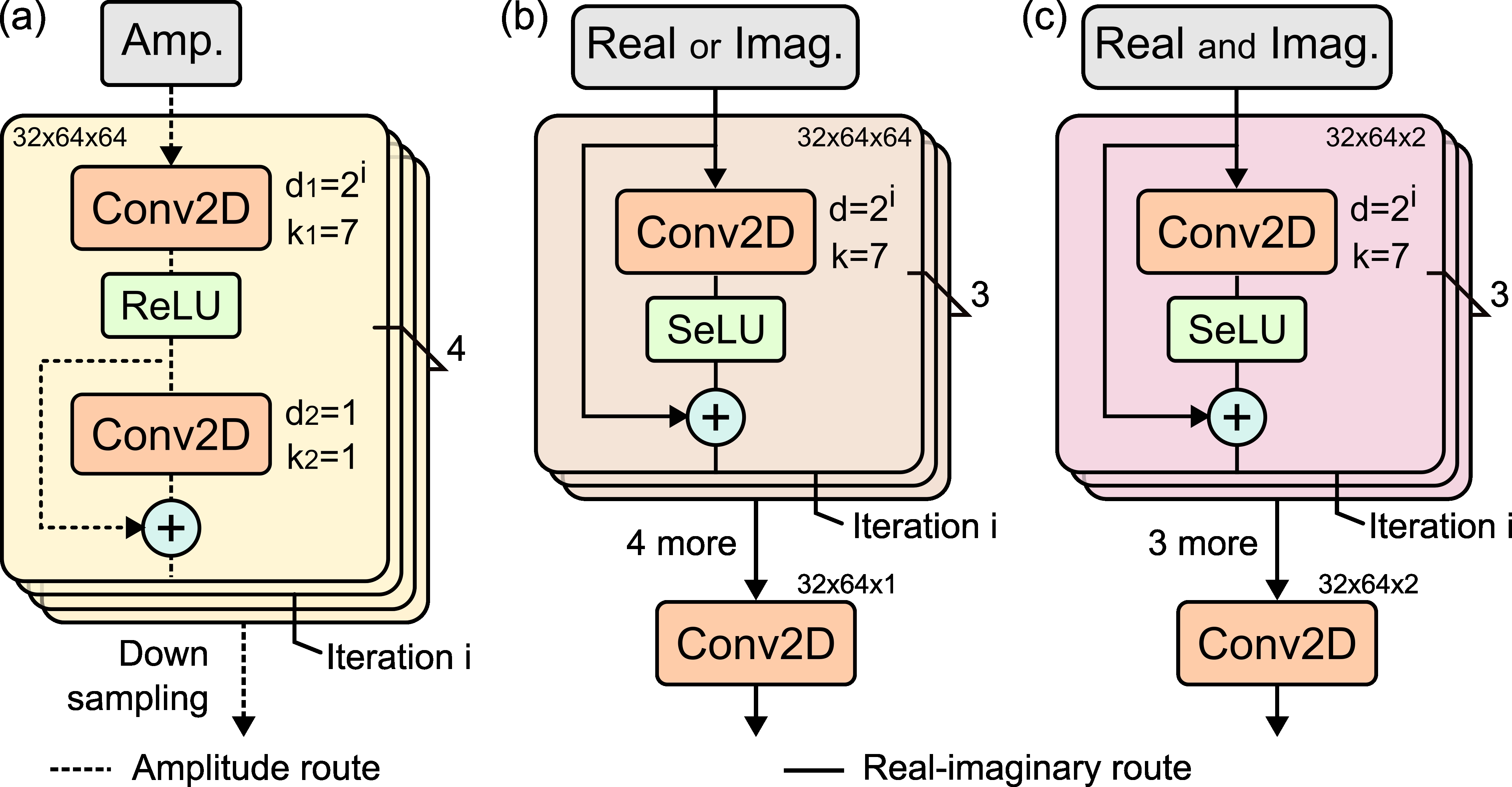}
    \caption{Detailed structure of the denoising deep learning model, specifically for (a) the first CNNs block of the amplitude autoencoder, (b) the phase alignment block, and (c) the cross-adjustment block. In each panel, d and k represent the dilation rate and the kernel size of the two-dimensional CNNs layers ({\tt Conv2D}), respectively. For the dilation rates, i $\geq0$ labels the current iteration. The sizes of {\tt Conv2D}s are shown in the blocks or next to the corresponding layers.}
    \label{fig:app_model_detail}
\end{figure}

\appendix
\section{Relation between scaling ratio, SNR, and overlap}
\label{app_rat_snr}
When the data set is created for the deep learning training initially, it is hard to directly decide the SNR. 
Not only the magnitude of the waveform, which will be determined by the scaling ratio, but also the current noise segment affect the SNR value. 
This can be seen by the outliers in Figure~\ref{fig:rat_snr_olp}, where a small scaling ratio can result in a large SNR. 
Although the relation may not be a 1-1 function, the SNR generally increases with the scaling ratio (Figure~\ref{fig:rat_snr_olp}). 
In addition, we plot the overlap between the denoised and the target waveforms in Figure~\ref{fig:rat_snr_olp}, where the color gradient provides a clear intuition that the model performs better for a larger SNR.

\section{Details of the denoising deep learning model}
\label{app_model_detail}
For the denoising deep learning model, we adopt CNNs layers to be the general backbone, but customize the design for each sub-structure. 
Given that the amplitude autoencoder deals with the firsthand noisy data, it consists of the deepest processing blocks among the model (e.g., Figure~\ref{fig:app_model_detail}(a)). 
For example, the data is passed through four iterations in the first CNNs block of the amplitude autoencoder, each of which consists of a stack of two-dimensional CNNs layers ({\tt Conv2D} in Figure~\ref{fig:app_model_detail}), {\tt ReLU} activation function, and skip connection. 
There are 64 channels to store the information of the amplitude spectrogram with the dimension $32\times64$ in different aspects.
For the first {\tt Conv2D} in each iteration, the square kernel with a side length of 7 is dilated by the powers of 2 (i.e., from 1, 2, 4 to 8) to capture the distant features. The second {\tt Conv2D} stands for a buffer, and its dilation rate is fixed. 
After the iterations, the data will be down-sampled to narrow the information flow, forcing the model to filter only the necessary and general characteristics. 
Such structure of the CNNs block is repeatedly seen in the rest of the amplitude autoencoder (Figure~\ref{fig:architecture}(a)). 

Since the phase recovery is simplified with the amplitude mask, as discussed in Section~\ref{sec:deep} and Section~\ref{subsec:flow}, the phase alignment and the cross-adjustment blocks of the Griffin-Lim network are operated in a more conservative manner (Figure~\ref{fig:app_model_detail}(b-c)).
To be specific, we assign a skip connection after every {\tt SeLU} activation function to propagate the neighboring features and to prevent any large changes.
The dilation rates of both blocks are also increased with the powers of 2 for the three iterations (i.e., from 1, 2 to 4). 
The phase alignment block (Figure~\ref{fig:app_model_detail}(b)) is deeper than the cross-adjustment block (Figure~\ref{fig:app_model_detail}(c)) by one more iteration group.
Another difference between these two sub-structures is about their input. 
The phase alignment block separately processes the real and the imaginary parts by treating them as individual spectrograms, whereas the cross-adjustment block regards the complex-valued spectrogram as one set of data with two channels (similar to RGB images). 
As a result, the {\tt Conv2D}s in the phase alignment block have shapes similar to those in the amplitude autoencoder. 
On the other hand, the filter sizes in the cross-adjustment block are fixed to be 2, one for the real part and the other for the imaginary part. 
By utilizing the phase alignment and the cross-adjustment blocks, the Griffin-Lim network (Figure~\ref{fig:architecture}(b)) thus is able to denoise the phase and finally returns a clean waveform.



\bibliographystyle{elsarticle-num} 
\bibliography{ref}






\end{document}